\documentclass[12pt,preprint]{aastex} 

\begin{document}
\newcommand{\pupa}{Puppis~A}
\title{Chandra X-ray Observation of a Mature Cloud-Shock Interaction in the Bright Eastern Knot Region of \pupa}
\author{Una Hwang\altaffilmark{1,2}, Kathryn A. Flanagan\altaffilmark{3}, Robert Petre\altaffilmark{1}}
\altaffiltext{1}{NASA Goddard Space Flight Center, Code 662, Greenbelt MD 20771; hwang@milkyway.gsfc.nasa.gov}
\altaffiltext{2}{Department of Physics and Astronomy, The Johns Hopkins University, 3400 Charles St, Baltimore MD 21218}
\altaffiltext{3}{Kavli Institute for Astrophysics and Space Research, MIT, 77 Massachussetts Ave, Cambridge MA 02139}

\begin{abstract}

We present Chandra X-ray images and spectra of the most prominent
cloud-shock interaction region in the Puppis A supernova remnant.  The
Bright Eastern Knot (BEK) has two main morphological components: (1) a
bright compact knot that lies directly behind the apex of an
indentation in the eastern X-ray boundary and (2) lying $1'$ westward
behind the shock, a curved vertical structure (bar) that is separated
from a smaller bright cloud (cap) by faint diffuse emission.  Based on
hardness images and spectra, we identify the bar and cap as a single
shocked interstellar cloud.  Its morphology strongly resembles the
``voided sphere'' structures seen at late times in Klein et al.'s
experimental simulations of cloud-shock interactions, when the
crushing of the cloud by shear instabilities is well underway.  We
infer an interaction time of roughly 3 cloud-crushing timescales,
which translates to 2000-4000 years, based on the X-ray temperature,
physical size, and estimated expansion of the shocked cloud.  This is
the first X-ray identified example of a cloud-shock interaction in
this advanced phase.  Closer to the shock front, the X-ray emission of
the compact knot in the eastern part of the BEK region implies a
recent interaction with relatively denser gas, some of which lies in
front of the remnant.  The complex spatial relationship of the X-ray
emission of the compact knot to optical [O III] emission suggests that
there are multiple cloud interactions occurring along the line of
sight.

%The
%Northern Knot region, by contrast to the Bright Eastern Knot, shows
%relatively little spectral structure.  It is located in a more
%isolated environment and is probably being viewed face-on, whereas the
%interaction at the Bright Eastern Knot is more complex, and is viewed
%edgewise.

\end{abstract}

\section{Introduction}

\pupa\ is a middle-aged member of the class of oxygen-rich remnants
that includes Cassiopeia A.  The optical emission from these remnants
is dominated by oxygen, requiring massive progenitors that synthesized
the oxygen during their hydrostatic evolution.  \pupa\ has faint
oxygen-dominated filaments (Winkler \& Kirshner 1985), whose proper
motions and positions give a dynamical age of 3700$\pm$300 yr for the
remnant (Winkler et al. 1988).  Oxygen ejecta also leave their imprint
on the X-ray line emission in high spectral resolution data from the
Einstein Observatory; the implied presence of at least 3 M$_\sun$ of O
and Ne ejecta requires a progenitor with an initial mass exceeding 25
M$_\sun$ (Canizares \& Winkler 1981).  A massive progenitor for \pupa\
is confirmed by the discovery of an isolated X-ray emitting neutron
star associated with the remnant (Petre et al. 1996), with no
confirmed period (Pavlov et al. 1999, 2002) and no radio counterpart
(Gaensler et al. 2000).

Based on the properties of the fast O-rich, and slower N-rich, optical
knots, Chevalier (2005) concludes that the progenitor of \pupa\
underwent substantial mass loss before exploding as a SN IIL/b.  The
red supergiant wind was likely to have been clumpy, and is expected to
have reached about 7 pc from the explosion center, in comparison to
the full 32 pc extent of \pupa.  This is consistent with the presence
of optically identified N-rich wind material in the interior of the
remnant, and indicates that the majority of the material with which
the remnant has interacted is interstellar rather than circumstellar.
Earlier in the remnant's evolution, the clumpiness of the
circumstellar environment may have also aided in the formation of
ejecta bullets by Rayleigh-Taylor instabilities (Jun et al. 1996).

Since massive stars typically evolve in associations and modify their
surroundings through their stellar winds, it is not surprising that
\pupa\ sits in an exceptionally complex interstellar environment.
Indeed, its X-ray properties are now largely shaped by its interaction
with its surroundings.  The remnant shows increasing X-ray surface
brightness west to east that reflects a large-scale increase in the
ambient density of roughly a factor of four (Petre et al. 1982).

The brightest X-ray feature is dubbed the Bright Eastern Knot (BEK)
for its location at the eastern edge of the shock front, and is
understood to arise from interaction of the shock wave with an
interstellar cloud.  The BEK also dominates the remnant at radio
wavelengths (Dubner et al. 1991), and its high radio brightness is
consistent with the cloud-shock interaction scenario.  The high
magnetic fields needed for particle acceleration to produce the radio
emission can be generated in vortices that are formed as a cloud is
disrupted by the shock (Stone \& Norman 1992).  In the northern
interior of the remnant, the Northern Knot (NK) is a more isolated
example of a shocked interstellar cloud.

The presence of an extremely massive and extensive atomic/molecular
cloud along the linear northeastern edge of \pupa\ (in the direction
towards the Galactic Plane), is established by observations of HI and
CO emission (Dubner \& Arnal 1988, Reynoso et al. 1995). Infrared
observations further reveal that the tip of the eastern molecular
cloud is probably projected in front of the remnant's eastern border
(Arendt et al. 1990).  Other massive HI and CO clouds to the southeast
and north are also likely to be interacting with the remnant (Reynoso
et al. 1995).  Indeed, the entire remnant sits at the periphery of a
very large HI shell of roughly 120 pc diameter and 30,000 M$_\odot$
estimated mass (Dubner \& Arnal 1988).  In this context, the
Bright Eastern Knot is just one part of an extremely complex and
extensive interstellar environment.

In this paper, we present Chandra images and spectra of the most
prominent shocked cloud interaction region in Puppis A---the Bright
Eastern Knot.
% and the more isolated Northern Knot.  
Cloud-shock interactions have been studied most extensively in X-ray
and optical studies of the Cygnus Loop (Levenson et al. 2002, 2005,
Leahy 2004, Patnaude et al. 2002, Graham et al. 1995, Fesen et
al. 1992).  The Cygnus Loop is expanding into a cavity wall, which is
itself a large expanding cloud (Leahy 2003).  The references above
identify a number of cloud-shock interactions in various parts of the
remnant, most of which are relatively recent.  They are
manifested by soft emission produced behind the decelerated shock
propagating into the cloud and enhanced harder X-ray emission produced
behind shocks reflected back into already-shocked gas.  The current
location of the blast wave in the intercloud gas is established by the
nonradiative optical emission.  It will be seen that our study of
\pupa\ is complementary to those of the Cygnus Loop in two ways: the
$\sim 700$ km/s intercloud shock velocities are higher than the
300-400 km/s velocities typical of the Cygnus Loop, and at least one
identified cloud-shock interaction in \pupa\ is significantly more
mature than those seen so far in the Cygnus Loop.

\section{Chandra Images and Spectra}
\subsection{X-ray Color Images}

We observed the Bright Eastern Knot in \pupa\ using the Chandra
Observatory's Advanced CCD Imaging Spectrometer (ACIS) for 10.6 ks,
%and 20.7 ks, respectively.  We also had an observation of a
%portion of the straight northeastern shock front, but that observation
%was marred by contamination from hard background flares (energies $>$
%2 keV) and is scheduled for re-observation.  
using only the backside-illuminated S3 chip to avoid telemetry
saturation in the event of strong background flares.  The mirrors
provide images with 0.5$''$ angular resolution over the $8' \times 8'$
single-chip field of view.  The CCD was operated in
%standard Faint mode for the NK, and
%the events file corrected to remove the effects of charge transfer
%inefficiency (CTI).  The BEK field is bright enough to warrant 
Graded mode, wherein the CCD events are characterized onboard the
spacecraft to reduce the telemetry load; the events thus could not be
corrected for charge transfer inefficiency.  The count rate was
relatively stable throughout the observation.
% for the BEK and NK observations.  

We supplement the Chandra images with lower resolution, but larger
field of view images from XMM-Newton and ROSAT.  We use the XMM-Newton
EPIC PN image from a 9.7 ks exposure, which provides a view of the
$\sim30'$ region near the BEK with a 6$''$ FWHM point-source image
core.  The XMM-Newton EPIC CCDs provide similar spectral resolution as
the Chandra ACIS instruments\footnote{The EPIC MOS camera data are not
useful for stuyding the region around the BEK because they were
collected with the central chip in partial window mode in order to
avoid pulse pile-up.  The high count rates also cause strong telemetry
dropout that is evident in the light curves from all the EPIC
detectors.}.  The ROSAT PSPC imaged the entire remnant in three
pointings of 2.3 to 7.9 ks exposure with 25$''$ FWHM angular
resolution and limited spectral resolution over its 0.2-2 keV
bandwidth.  We reduced the PSPC data using the algorithms and software
described in Snowden et al. (1994) to model and subtract the
background.  The EPIC and PSPC data are used here for spectral imaging
only, and not for detailed spectral analysis.

The Chandra observation is shown in a three color rendering in Figure
1 (soft energies 0.4-0.7 keV, covering O emission and continuum, in
red; medium energies 0.7-1.2 keV, covering Ne and Fe L, in green; hard
energies $>$1.2 keV, in blue).  Intricate filamentary structure is
evident in the image field.  The indentation seen in the eastern X-ray
boundary strongly suggests that the shock wave is wrapping around an
obstacle of approximately 40$''$ angular extent on the sky (0.4 pc at
a distance of 2.2 kpc for \pupa: Reynoso et al. 2003, 1995, Winkler et
al. 1988, Dubner \& Arnal 1988; though Woermann et al. 2000 suggest a
smaller distance). A compact, X-ray bright region is located
immediately west of this indentation.  Roughly $1'$ further west lies
a more extended vertical structure (henceforth the bar) that bends
sharply inward at its bright southern end, has a sharp boundary to the
east and a more diffuse boundary to the west.  The bar is separated by
lower surface-brightness emission from a smaller cloud (cap) lying
$4'$ away near the western edge of the Chandra field.  The bar appears
to separate the relatively soft spectra found to the east from harder
spectra to the west.

The EPIC PN spectral images shown in Figure 2 use the same energy cuts
as the Chandra ACIS spectral images, whereas the ROSAT PSPC mosiacs of
the entire remnant use the three standard PSPC energy bands (Snowden
et al. 1994; soft bands 1L and 2 covering energies 0.11-0.28 keV in
red, medium bands 4 and 5 covering 0.44-1.21 keV in green, and hard
bands 6 and 7 covering 0.73-2.04 keV in blue).  These images verify
that the bar and cap behind the shock front indentation are spectrally
similar.  They also show that a large region to the south of the BEK
is spectrally soft, whereas the regions to the north of the BEK and
running behind the entire straight northeastern shock front are
spectrally hard.  This large-scale band of hard emission has been
previously noted from ROSAT All-Sky Survey and ASCA observations
(Aschenbach 1994, Tamura 1995). Just south of this hard region,
relatively faint emission from the nearby Vela supernova remnant is
seen extending beyond \pupa\ to the east and southwest in the PSPC
mosaic, but is appreciable only in the softest image because of Vela's
extremely soft spectrum (see Lu \& Aschenbach 2000). It is likely that
this entire Vela filament is projected in front of Puppis A.  As in
the smaller Chandra images, filamentation is observed throughout
the remnant (and is seen to even better advantage in the ROSAT High
Resolution Imager mosaic of Petre et al. 1996).
 
%The NK field has relatively little spatial structure in the spectral
%hardness map, except that the knot itself is slightly harder than its
%surroundings.  The knot is bright and compact, and forms a corner
%on its western side. It is surrounded by fainter patchy emission.

\subsection{Spectral Maps}

Since spectral hardness may be affected not only by temperature, but
also by column density, relative element abundances, and the
ionization state of the X-ray emitting gas, we examined the X-ray
spectra across the ACIS field.  The source region on the CCD was
divided into a grid of continuous regions for spectral analysis,
starting with a basic box-shaped region that was further subdivided
according to the number of counts contained.  The 250 spectral regions
range from 12.5$''$ to 50$''$ boxes or rectangles.
%and the 148 regions for the NK from 17.5$''$ to 70$''$.  
Spectra typically contained 7000-12000 counts, though the actual
number of counts ranged from 4000 to 16000.  Although we do not use
the EPIC data for spectral analysis here---there are differences in
point-spread function that would make a detailed comparison with the
Chandra data difficult---we have verified that the general {\it
qualitative} features of the ACIS spectral results are also seen in
the EPIC data.

To fit the spectra, we used simple models that account for the
nonequilibrium ionization (NEI) state of the gas, where the ionization
state is characterized by an ionization timescale or ``age'', defined
as the product $n_et$ of electron density $n_e$ and time $t$ since
shock-heating of the gas.  Specifically we use the $pshock$ model in
XSPEC, which assumes a single temperature and a range of ionization
ages, which we take from zero up to a fitted maximum value.  This
should better describe the complex spectra in extended spatial regions
than would a single-temperature, single-ionization age NEI model, and
we generally find that to be the case for these data.  Solar
abundances from Anders \& Grevesse (1989) are used throughout.  To
model the buildup of contaminants on the detector surface, we include
the ACISABS model in our spectral fits, specifying the number of days
since launch for the observation (833).
% 958 for NK).
To subtract the background, we use local off-source regions.
% for the
%BEK.  Since the entire NK field is covered by remnant emission, we
%take its background spectrum from the shock front observation, which
%was also taken in FAINT mode, rather than from the BEK observation,
%which was taken in GRADED mode.  Even with the worst hard flares
%removed from the shock front observation, this is an over-subtraction
%of the high energy background for the NK spectra, so we partly
%compensate for this by fitting the spectra only up to 4 keV.
%(*use another field?)

Figure 3 shows the spectral grid regions and image maps for the fitted
temperatures, ionization ages, and column densities. The BEK region
shows significant spectral complexity, with low temperatures kT of
$\sim$0.3 keV near the eastern indentation, to 0.5-0.6 keV in the
regions that cover the bar and cap, all the way up to high
temperatures above 0.8 keV in some low surface brightness regions.
For the four representative regions indicated in Figure 3, the spectra
and fitted models are shown in Figure 4, and the fitted parameters
given in the Table.  The errors cited in the Table and discussed in
the paper are the 90\% confidence ranges for a single interesting
parameter, corresponding to $\Delta\chi^2$= 2.71.

There are two important features to note in the spectral parameter
maps of the BEK.  First, spectra in the bar and cap (as well as most
of the associated lower surface brightness regions) are seen to be
characterized by similar temperature, maximum ionization age, and
column density.  Second, the regions to the east and far southeast
(i.e., the lower left region of the ACIS field) of the bar are
spectrally very distinct: their temperatures are lower (at $\sim$0.3
keV compared to $\sim$0.6 keV for the bar and cap), ionization ages
lower (below 10$^{11}$ cm$^{-3}$s compared to several 10$^{11}$
cm$^{-3}$ s), and column densities higher (3-5 $\times 10^{21}$
cm$^{-2}$ compared to 1-2$\times 10^{21}$ cm$^{-2}$).

A closer look at the temperature change from the compact knot to the bar is
given in Figure 5.  The spectra in a series of adjacent rectangular
regions were fitted with single-temperature parallel shock models
assuming solar abundances to obtain the temperature, ionization age,
and column density.  The temperature change across the series is
gradual and smooth across the first four regions, rather than being
abrupt, but the changes are more abrupt for the other parameters: the
column density drops sharply for the fourth region, just as the
ionization age jumps up.

For spectra with at least 6000-7000 counts, typical errors are a few
to 10\% for the fitted temperature, and roughly 20\% for the column
density and ionization age.  In the cases where the spectra contain
fewer counts, these errors are correspondingly larger, up to 50-60\%
or higher for the column density and ionization age.  The temperature,
however, is determined to 10\% or better in nearly all cases.  For
most of the spectra, there is a clear-cut global minimum, but in a
number of more ambiguous cases, there are two comparable local
minima---usually one with very low column density and high
temperature, and the other with high column density and low
temperature.  This accounts for a few isolated cool (or hot) regions
in the temperature map in Figure 3.  The column densities fitted for
some regions are unrealistically low for \pupa\ ($<10^{20}$
cm$^{-2}$), and most likely indicate that the simplified model
assumptions imposed for the spectral fits are inadequate.

We take $\chi^2_r=\chi^2 /\nu$ as the goodness-of-fit criterion, where
$\nu$ is the number of degrees of freedom.  These values are shown in
Figure 6.  In general, they are less than 2, which we loosely take to
be a reasonably acceptable fit in view of the limitations of the
spectral model and systematic effects in the data.  Some 30\% of the
spectra fitted have $\chi^2/\nu$ greater than 2, however, and as these
are not necessarily the spectra with the highest signal-to-noise, the
poorness of fit must be in part due to the shortcomings of the
spectral model.  We tried a few more complex models to try to improve
these fits, and found the largest improvements came from allowing the
overall O-Ne element abundances to be fitted independently from the
others (while tied in their solar abundance ratio). An abundance
enhancement for these two elements of 2-3 times their solar value
greatly improves the fits in a subset of the spectra that initially
had poor fits.  An example is given in the table (region 214).  Where
the improvement in $\chi^2$ is statistically significant (we took a
$\Delta\chi^2/\chi^2_r$ value of 7.88 as the cutoff), we replaced the
original fit with the revised fit, as shown in Figure 6.  In the end,
roughly one-fifth of all the regions remain relatively poorly fitted
with $\chi^2/\nu > 2$.

In the revised fits, the abundance enhancements generally occur in
localized pockets.  For the most part these are outside the bar and
cap---to the north, southeast and east (north of the compact knot).
field.  In about half of the cases where the enhancements occur, the
fitted temperature does change, usually from $\sim0.6$ keV down to a
lower value, with relatively high ionization ages that are well above
$10^{11}$ cm$^{-3}$s (note that the cool regions near the compact knot
have lower ionization ages below $10^{11}$ cm$^{-3}$s).  In the
following section, we will interpret the bar and cap as components of
a single, shocked interstellar cloud.  As would be expected in this
case, the dense cloud material is not as contaminated by ejecta than
is less dense intercloud gas.  While the shocked cloud is slightly
less coherent in the revised temperature map, it remains readily
identifiable.
% 5-17-05: 117 regions with signifcant change in kT and chi^2; 
% 64 with more than 0.2 keV difference in kT; 
% 38 of these from kT > 0.55 to kT < 0.4

Enhanced element abundances are not generally expected in older
remnants, but in \pupa\ they are in line with the X-ray results of
Canizares \& Winkler (1981), the optical identification of ejecta by
Winkler \& Kirshner (1985), and the expectation that the formation of
ejecta clumps was aided by Rayleigh-Taylor instabilities arising from
the interaction of the shock with a clumpy circumstellar and
interstellar environment.  Furthermore, the BEK region is just a few
arcminutes beyond the location of a number of optically-emitting O
ejecta knots, which are preferentially located toward the east (see
Winkler et al. 1988).  Abundance enhancements have also been noted in
the Cygnus Loop.  Those found in the center of the Cygnus Loop remnant
have been interpreted as being due to enrichment by ejecta (Miyata et
al. 1998), whereas the enhancements in the southwest have been
suggested to be the result of enrichment of the ISM by an earlier
supernova explosion (Leahy 2004).

%In contrast to the BEK, the spectral parameters are fairly uniform
%throughout the NK field.  The fitted temperature is low at $kT
%\sim0.3-0.4$ keV, the ionization ages fairly high at above $10^{11}$
%cm$^{-3}$s, and the column densities also high at 4-5 $\times 10^{21}$
%cm$^{-2}$.  There is seen to be no significant difference between the
%brighter filaments and the fainter diffuse regions, though the bright
%core of the knot is slightly harder than the surroundings, and does
%have a slightly higher temperature.  The maps for the NK are shown in
%Figure 5.  In general these fits are not of high quality, with only a
%third or so of the fitted regions having $\chi^2/\nu$ below 2.  This
%is due to a combination of imperfect background subtraction, and the
%spectra requiring a more complex model than a simple,
%single-temperature plane-parallel shock.  This may be consistent with
%a face-on view of the NK, but it is difficult to disentangle this from
%the background subtraction problem.

\section{Discussion}

\subsection{Cloud-Shock Interaction}

The X-ray images and spectra of the bar and cap in \pupa\ suggest that
they may be components of a single physical structure.  We interpret
this composite structure to be an interstellar cloud that has been
shocked, and is in the process of being disrupted by hydrodynamic
instabilities.

The basis of our premise is that the morphology of the bar and cap
strongly resembles those seen in experimental simulations of the
interaction of supernova shock waves with dense interstellar clouds by
Klein et al. (2003). Taken together, the bar and cap suggest a
spherical shell whose hemispheres have been pulled apart---a ``voided
sphere'', as described by Klein et al. and pictured in their Figure
15.  Though the cloud-shock interaction occurs over hundreds of years
on parsec scales in supernova remnants, its morphology and evolution
can be simulated in laboratory experiments of some 20 orders of
magnitude smaller spatial and temporal scale by virtue of the
self-similarity of the hydrodynamics under certain transformations
(see Klein et al. 2003 and Ryutov et al. 1999 for details).  The
experiments involve a dense 100 $\mu$m copper ball representing the
interstellar cloud embedded in low-density plastic representing the
intercloud medium, with a density contrast of roughly a factor of ten.
A strong Mach 10 shock wave is initiated in the plastic by
laser-produced X-rays that ablate the plastic.  The copper sphere is
vaporized before arrival of the shock wave by heating from X-rays
coming from the gold capsule encasing the system.  The progress of the
interaction is recorded edge-on by doing the experiment in multiple
sets and using an X-ray backlight to make shadowgrams at different
time delays relative to the beginning of the laser pulse.

As described by Klein et al., the characteristic time scale of the
interaction is the cloud-crushing time $t_{cc}$, which is defined as
$a_o/v_s$, where $a_o$ is the initial radius of the cloud and $v_s$ is
the velocity of the shock in the cloud.  For a density enhancement of
$\chi$ for the cloud relative to the intercloud medium, the velocity
of the blast wave in the intercloud medium $v_b$ is decreased by a
factor of $\chi^{1/2}$ within the cloud, so that $v_s=v_b/\chi^{1/2}$.
For the Klein et al. experimental configuration, the cloud-crushing
time is 8 ns, and the experiment follows the progress of the
cloud-shock interaction for several $t_{cc}$.  For collisionless
strong shocks where conduction, radiation, and viscosity may
reasonably be ignored, $t_{cc}$ is the single relevant parameter for
describing the evolution of the system.

The experimental results are documented in detail in Figures 14 and 15
of Klein et al. As the shock traverses the cloud, the cloud is
initially compressed along the direction of the shock velocity, then
begins to expand in the orthogonal direction because the shocks are
weaker at the sides than along the front and back of the cloud.  The
cloud takes on an umbrella-like shape as arms begin to form at the
sides of the cloud from the action of shear instabilities.  As the
cloud expands into the intercloud medium, and hydrodynamical shear
flow instabilities take their toll, the front and back faces of the
cloud begin to separate from each other at a time of roughly 3
$t_{cc}$ (see the t=43.6 ns radiogram in Klein et al.).  This is the
``voided sphere'' morphology that is displayed by the bar and cap
structure in \pupa.  We believe this is the first instance in which
such a structure has been identified astrophysically.  The cloud-shock
interactions studied in the Cygnus Loop (see references cited in the
introduction) are all in an earlier phase ($\sim t_{cc}$), which is
too early for development of the voided sphere morphology.

The voided sphere structure is confirmed in three-dimensional
hydrodynamical simulations of the interaction by Klein et al. (their
Figure 19), but interestingly, is not manifested in two-dimensional
simulations.  Evidently, the third dimension is critical to the
development of the instability that shapes the cloud at late times.
The simulations also show that the face of the cloud closest to the
shock front (in our case, this would be the eastern edge of the bar)
has an irregular and fluted outer surface, though this is not
particularly apparent in either our X-ray images or the experimental
radiograms.

To translate the interaction time for the cloud in \pupa\ into
approximate years, we consider the size and temperature of the cloud.
For a distance of 2.2 kpc, the angular size of the cloud, which is
about 4$'$ east-west and 4.7$'$ north-south, translates to 2.5 by 3.0
pc.  We take the current average ``radius'' of the cloud to be about
1.4 pc.  By visual examination of the experimental radiograms, the
size of the cloud at time 3.7 $t_{cc}$ (t=43.6 ns), is about 1.3 times
its original axial dimension along the shock velocity and 2.7 times
its original dimension in the orthogonal direction.  We thus take the
average expansion of the cloud at this epoch to be factor of 2,
keeping in mind a roughly 30\% error, and the possibility that the
actual amount of compression might vary depending, for example, on the
density enhancement of the cloud.  We thus estimate the original cloud
radius to be about 0.7 pc, or $2.2\times 10^{13}$ km.  The temperature
of the X-ray emitting gas associated with the cloud is 0.6 keV, which
corresponds to a shock velocity of 700 km/s if we assume there is full
equipartition of the shock energy between electrons and ions.  The
cloud crushing time may then be estimated as $a_o/v_s$ = 980 years,
give or take some 30\%.  Taking the epoch of the interaction to be 3.7
$t_{cc}$, this translates to an interaction age of 2000-4000 years.
This is a significant fraction of the estimated remnant age of
3700$\pm$300 (Winkler et al. 1988), and even exceeds it at its upper
end, so we take the lower end of the interaction age range as being
more plausible.
%The pulsar characteristic age is somewhat higher at 8000 years (Pavlov et
%al. 1999), but is not a reliable age indicator (e.g., Gaensler \&
%Frail 2000, Migliazzo et al. 2000).  

In fact, a 3700 yr age for the remnant may seem low given that it
implies rather high average expansion velocities of 5000 km/s for the
western limb (for the expansion center determined by Winkler et
al. 1988), but this dynamical age should be an upper limit unless the
optical knots have somehow been accelerated rather than decelerated.
Somewhat lower average expansion velocities of 3000 km/s are
determined for the eastern limb, as might be expected given the
significant and widespread deceleration of the shock by large
molecular and interstellar clouds.

As an order-of-magnitude check on our inferred interaction time, we
may use the distance of the shocked cloud from the optical expansion
center determined by Winkler et al. (1988).  The ratio of the radial
distance from the expansion center to the shocked cloud compared to
the radial distance to the farthest region on the western limb is
about 0.55.  We can estimate that the shock would have reached the
location of the cloud in 55\% of its 3700 yr age, or 2000 yrs, if we
assume for this purpose that there has been minimal deceleration to
the western tip of the remnant, and that this cloud is the first major
obstacle encountered by the shock in the east.  Then the shock time of
the cloud was 1700 years ago, which is in relatively good agreement
with the lower end of the range of interaction times inferred from the
morphology and spectrum of the cloud.

In reality, the initial shock-cloud interaction could actually have
occurred somewhat earlier than our estimate.  First, we have probably
underestimated the shock velocity in the cloud by assuming full
equipartion of the ion and electron energy.  This is not
self-consistent, since only partial equipartition is achieved at the
shock velocity of 700 km/s that we have determine for the cloud
(Rakowski et al. 2003), and a higher shock velocity would decrease the
interaction time.  Second, the current location of the cloud is likely
to be farther from the (optically determined) expansion center than
its original location because the shock interaction accelerates the
cloud so that it is eventually co-moving with the shocked intercloud
gas (McKee, Cowie \& Ostriker 1978, Klein et al. 1994).  This
acceleration takes place on a timescale of a few $t_{cc}$, so that our
cloud should by now be moving with the intercloud gas.

The density of the shocked cloud may be obtained from the spectrally
fitted ionization age $n_et$, which is the product of electron density
and shock time, and the estimated shock-cloud interaction time.  The
fitted ionization age is the maximum value required in the models,
which include a range of ionization ages from zero up to the maximum.
From Figure 3, we take the typical maximum ionization age in the
shocked cloud region of the BEK to be $5\times 10^{11}$ cm$^{-3}$s.
We may then divide out an interaction time of 2000 yr, at the more
plausible lower end of the estimated range as discussed above.  To
relate the electron density in the shocked gas to the initial density
of the cloud, we first note that the gas in the shocked cloud is
compressed by a factor of 2 over its original density according to
Klein et al. (1994).  We can then relate the electron density $n_e$ in
the shocked cloud to the initial H density $n_{Ho}$ in the unshocked
cloud as $n_e=1.23 n_H = 2.46n_{H_0}$ for a solar abundance plasma, to
obtain a corresponding initial average H density in the cloud of 3
cm$^{-3}$.  An estimated factor of ten enhancement of the density in
the cloud (i.e., as in the experimental simulations) implies local
interstellar densities of 0.3 cm$^{-3}$.  The inferred local
interstellar density may be as low as 0.15 cm$^{-3}$ if we consider
the upper limit of 4000 yr estimated for the interaction age.  A $>3$
times higher density of 10 cm$^{-3}$ is obtained for the BEK region
from analysis of the Einstein High-Resolution X-ray images (Petre et
al. 1982), but this is for a higher-surface brightness region centered
on the more recently shocked compact knot that lies eastward of the
bar and cap.  All of these estimates are rather uncertain, but serve
to illustrate the possibility of diagnosing shock-cloud interactions
from a combination of X-ray observations and laboratory experiments.

%\begin{figure}
%\label{bek}
%\centerline{\includegraphics[scale=0.5]{klein.ps}}
%\figcaption{Radiograph of a laboratory simulation of a cloud-shock
%interaction from Klein et al. (2003).  Taken from their Figure 15 at
%t=43.6 ns, corresponding to 3.7 $t_{cc}$.}
%\end{figure}

\subsection{Projection Effects in the BEK}
 
The region of the BEK immediately behind the shock front is spectrally
distinct: the temperatures are low, ionization ages low, and the
column densities higher than elsewhere in the ACIS field.  The 0.3 keV
temperature determined for this region is supported by the presence of
exceptionally bright optical coronal [Fe XIV] $\lambda\ 5303$\AA\
emission (Teske \& Petre 1987).  This emission can arise in relatively
hot plasmas with a narrow range of temperatures between kT = 0.07-0.3
keV that overlaps the X-ray measured temperature at its upper end.  In
the western region, where the fitted X-ray temperatures are higher,
[Fe XIV] emission is not prominent.

The low temperatures in this region suggest slow shocks propagating
into relatively dense clouds.  A simple scaling of the X-ray emitting
temperatures suggests that the density of the compact cloud currently
being encountered by the shock is some 4 times higher than that of the
older shocked cloud discussed above, though this is obviously a crude
estimate that ignores the complications due to multiple shock
reflections.  A close look at the temperature behavior at the
interface between the $\sim$ 0.3 keV emission associated with the
compact knot and the hotter gas associated with the bar and cap shows
that the temperature changes gradually, presumably due to a
combination of projection effects and the deceleration of the shock
upon hitting the large eastern molecular cloud with which it is now
interacting.  The sharp increase in the column density is consistent
with the presence of dense foreground gas associated with these
clouds, as observed by IRAS (Arendt et al. 1990).  The relatively low
ionization ages determined for this portion of the BEK, combined with
the compact morphology of the knot at the head of the indentation,
suggests a relatively recent interaction, consistent with its position
at the eastern outer edge of the remnant.  The sharp jump in
ionization ages is somewhat puzzling.  One speculative possibility is
that the shock interacted primarily with diffuse intercloud gas for
some time after encountering the cloud undergoing the shear
instability before being strongly decelerated by the large molecular
cloud system with which it is now interacting.

Diffuse and filamentary [O III] emission are also present in this
region of the BEK.  The [O III] image from Blair et al. (1995) is
shown in Figure~7 with the Chandra X-ray image and contours
overlaid.\footnote{An overlay of this [O III] image with the lower
angular resolution image from the ROSAT HRI is shown in the paper by
Blair et al. 1995.} The optical coordinates were registered using HST
Guide Stars, but no effort was made to correct for proper motions,
which would in any case be slight.  As noted by Blair et al., the [O
III] emitting filaments have clearly been shocked, with a velocity of
160-180 km/s determined from the UV spectrum of one of the filaments.
At these velocities, the shock is clearly propagating into very dense
material.  A correlation between the optical and X-ray emission is not
expected, since the temperatures behind these shocks are too low to
produce X-ray emission.  It is notable, however, that the [O III] is
located significantly beyond the eastern boundary of the X-ray
emission, with the northern and southern edges of the indentation in
the X-ray emission emerging from within the optical boundaries.  The
[O III] would thus appear to indicate the easternmost extent of the
blast wave in this region, except that the X-ray morphology strongly
suggests a shock wrapping around an obstacle.  The pronounced
indentation of the X-ray emitting shock at the cloud indicates that
this obstacle is probably quite extended along the line of sight.  The
situation here suggests that projection effects are important, with
multiple clouds in the line of sight---a possibility previously noted
by Blair et al. (1995). It is certainly plausible that the dense
molecular gas with which the remnant is now interacting would be
associated with clouds with a spectrum of sizes and masses. Petre et
al. (1982), for example, estimate density inhomogeneities on spatial
scales ranging from 0.1 to 5 pc in the remnant.

\section{Summary and Conclusions}

Chandra images and spectra support that the \pupa\ supernova remnant
is interacting with a complex system of interstellar clouds.  In this
paper, we have identified a shocked cloud behind the eastern shock
front that is in a relatively late phase of evolution and is being
torn apart by shear instabilities.  This is the first astrophysical
example of a cloud-shock interaction in this mature stage of evolution
that has been identified in X-ray emission.  We diagnose the
evolutionary state of the interaction by direct comparison of its
morphology with scaled laboratory simulations of the cloud-shock
interaction.  Using the X-ray morphology of the cloud, together with
X-ray images and spectra, we estimate that the onset of interaction
was roughly 3 cloud-crushing times, or 2000-4000 years ago, and that
the initial cloud density was 1.5-3 cm$^{-3}$.  There are considerable
uncertainties inherent in the details of these estimates, including
the effects of electron-ion equipartition, the effect of secondary
shocks on the observed temperatures, the apparent contamination of the
diffuse gas by ejecta, and the relatively crude method of
correspondence to the experimental results.  The general
identification of a mature cloud-shock interaction, however, is quite
clear.

Closer to the forward shock in the BEK region, the shock has recently
interacted with a more dense and extended obstacle.  The spectral
characteristics of the front portion of the BEK region are uniquely
different from those in the rest of the Chandra field, with lower
temperatures supported by the presence of [Fe XIV] emission, lower
ionization ages, and higher column densities.  These inferred spectral
properties, and comparison of the X-ray/optical morphology, suggests
that the remnant is now interacting with multiple clouds in this
region, some of which are in the remnant foreground.

It is likely that numerous such cloud-shock interactions have left
their imprint throughout the remnant.  The intricate filamentary
structures that pervade the X-ray images of \pupa\ shown here and
elsewhere resemble the final outcome of numerical hydrodynamical
simulations that model the disruption of clouds as they are overrun by
shocks (Stone \& Norman 1992).  The shocked clouds studied here will
also ultimately contribute to this filamentary structure in \pupa.

\acknowledgments We thank Kristy Dyer and Martin Laming for
stimulating discussions, and Estela Reynoso for discussion of the
distance to \pupa.

\clearpage
\begin{deluxetable}{ccccccc}
\tabletypesize{\scriptsize}
\tablecaption{Sample Spectral Fits}
\tablewidth{0pt}
\tablehead{
\colhead{Region}&\colhead{Counts}&\colhead{$\chi^2, \chi^2/\nu$}&\colhead{N$_{\rm H}$ ($10^{22}$ cm$^{-2}$)} & \colhead{kT (keV)}&\colhead{$n_et$($10^{11}$ cm$^{-3}$s)}&\colhead{O/Ne (rel $\odot$)} \\
}
\startdata
127 (bar) & 8211 & 121.6, 1.74 & 0.059 (0.027-0.13) & 0.57 (0.55-0.59) & 7.7 (5.9-9.6) & ...\\
139 (cap) & 13700 & 129.0, 1.59 & 0.17 (0.14-0.21) & 0.58 (0.56-0.60) & 3.8 (3.0-4.7) & ... \\
74 (compact) & 9036 & 67.8, 1.17 & 0.33 (0.30-0.37) & 0.33 (0.29-0.37) & 0.49 (0.38-0.59) & ... \\
212 (hard) & 9878 & 106.7, 1.31 & 0.087 (0.050-0.12) & 0.76 (0.72-0.80) & 1.8 (1.4-2.2) & ... \\
%19 (hard) & 10028 & 119.7, 1.81 & 0.017 ($<$0.042) & 0.92 (0.85-1.0) & 0.82 (0.65-1.1) & ...\\
%19 (hard) & ... & 117.8, 1.81  & (0.023-0.18) & (0.87-0.96) & (0.72-0.93) & (1.0-1.2) & ... \\
214  & 9790 & 145.3, 1.94 & 0.13 (0.12-0.14) & 0.81 (0.78-0.84) & 1.3 (1.2-1.4) & ...\\
214  & ...  & 104.5, 1.41 & 0.40 (0.35-0.44) & 0.66 (0.58-0.74) & 0.75 (0.50-1.1) & 2.6 (2.3-3.0) \\
\enddata
\tablenotetext{*}{Errors quoted are 90\% confidence for one interesting parameter ($\Delta\chi^2=2.7$).}
\end{deluxetable}

\clearpage

\begin{figure}
\epsscale{0.4}
\label{fig-1}
\plotone{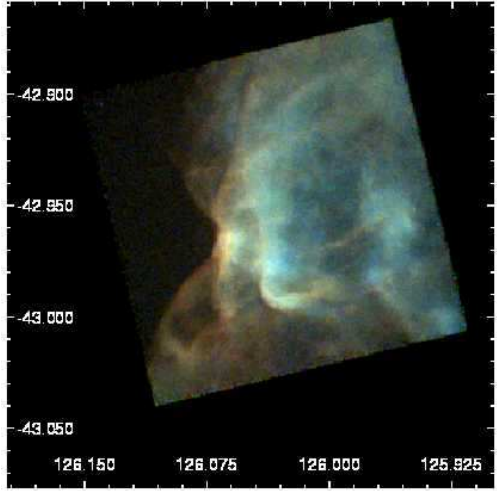}
\bigskip
\caption{Three-color Chandra images with soft band (O) in red, medium
band (Ne) in green, and hard band in blue of the BEK.  The intensity
scale is square-root, and J2000 coordinates are given with N pointing
up.}
\end{figure}

\begin{figure}
\epsscale{1.0}
\label{fig-2}
\plotone{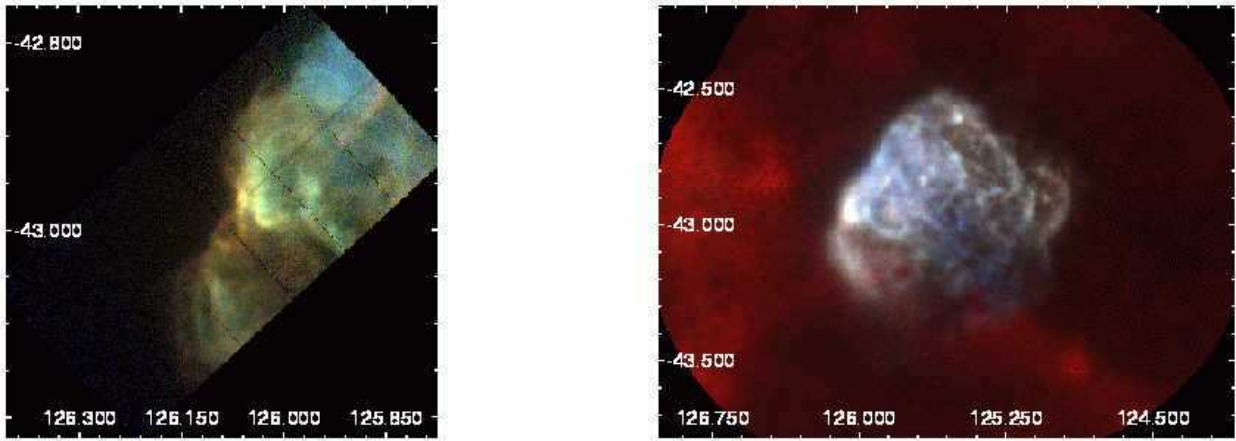}
\caption{(a) Three-color image of the larger angular scale region
around the BEK from the EPIC PN camera on XMM-Newton, using the same
energy cuts as for the Chandra images in Figure 1. (b) Adaptively
smoothed ROSAT PSPC mosaic of the entire remnant in the standard three
PSPC colors: soft in red, medium in green, hard in blue (see text and
Snowden et al. 1994).}
\end{figure}

\begin{figure}
\epsscale{1.0}
\label{fig-3}
\plotone{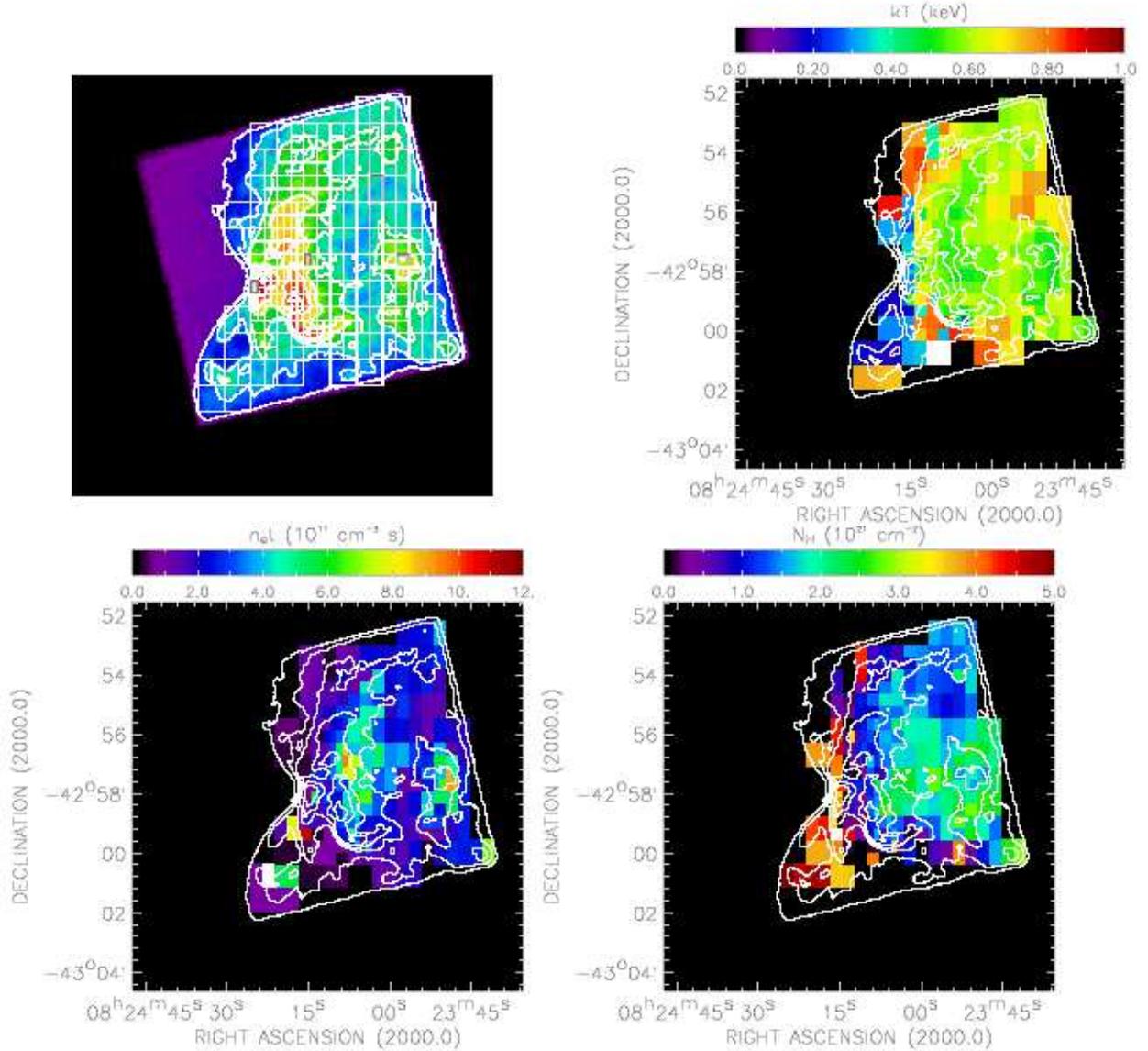}
\caption{(a) BEK spectral regions overlaid on a smoothed square-root
% 6 pixel sigma 
intensity image showing: a marked indentation of the X-ray emitting
blast wave to the east, the compact knot just west of the indentation
in red and white, the bar to the west of the compact knot running
roughly north-south in red and yellow, the smaller cap in yellow and
green near the western edge of the field.  The spectra of the regions
indicated in black are shown in Figure 4 (from left to right: compact
knot, bar, hard region at top right, and cap at lower right).  The same
smoothed intensity contours are overlaid in each panel of this
figure. (b) Temperature (kT) map in keV; (c)
Ionization age ($n_et$) map in $10^{11}$ cm$^{-3}$ s; (d)
Column density (N$_{\rm H}$) map in $10^{21}$ cm$^{-2}$.}
\end{figure}

\begin{figure}
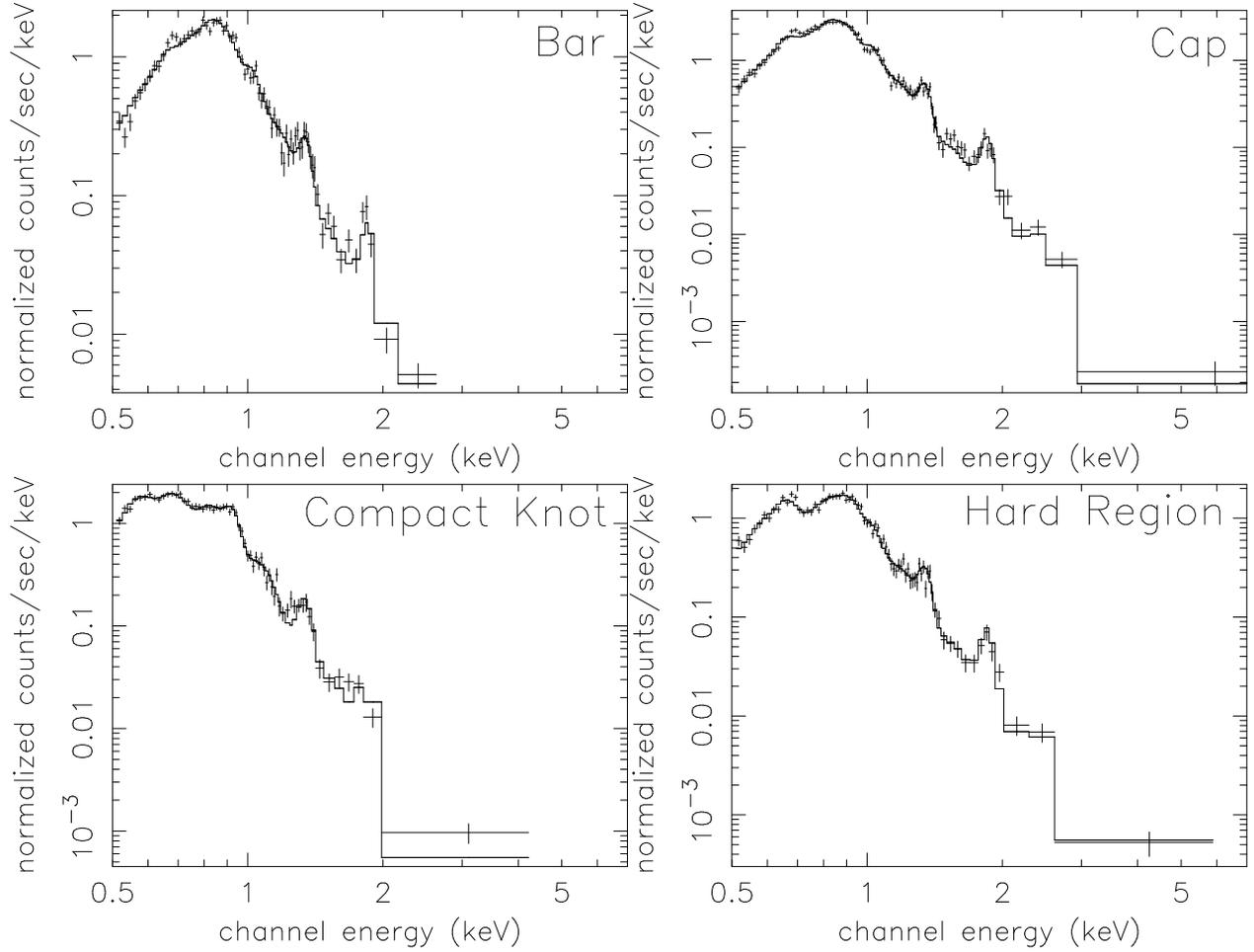

\epsscale{0.4}
\label{fig-4}
\centerline{\includegraphics[scale=0.4,angle=-90]{f4a.ps}\includegraphics[scale=0.4,angle=-90]{f4b.ps}}
\centerline{\includegraphics[scale=0.4,angle=-90]{f4c.ps}\includegraphics[scale=0.4,angle=-90]{f4d.ps}}
\caption{Representative spectra in the BEK field, taken from the
regions indicated in Figure 3: (a) from bar and (b) cap
structure; (c) from soft compact knot; (d) from a
hard region north of the bar and cap.  See the table for fitted
parameters.}
\end{figure}

\begin{figure}
\epsscale{0.4}
\label{fig-5}
\plotone{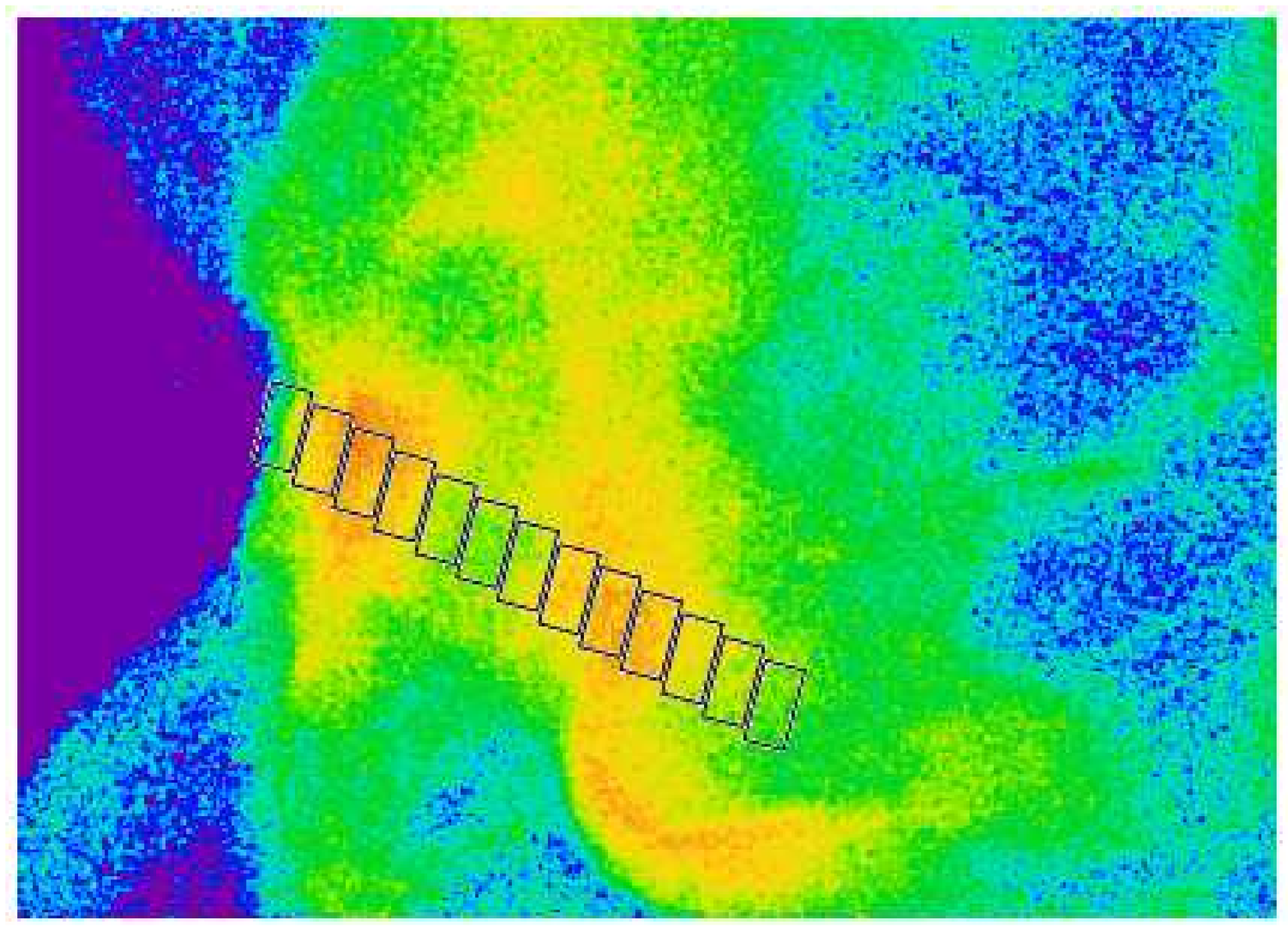}
\plotone{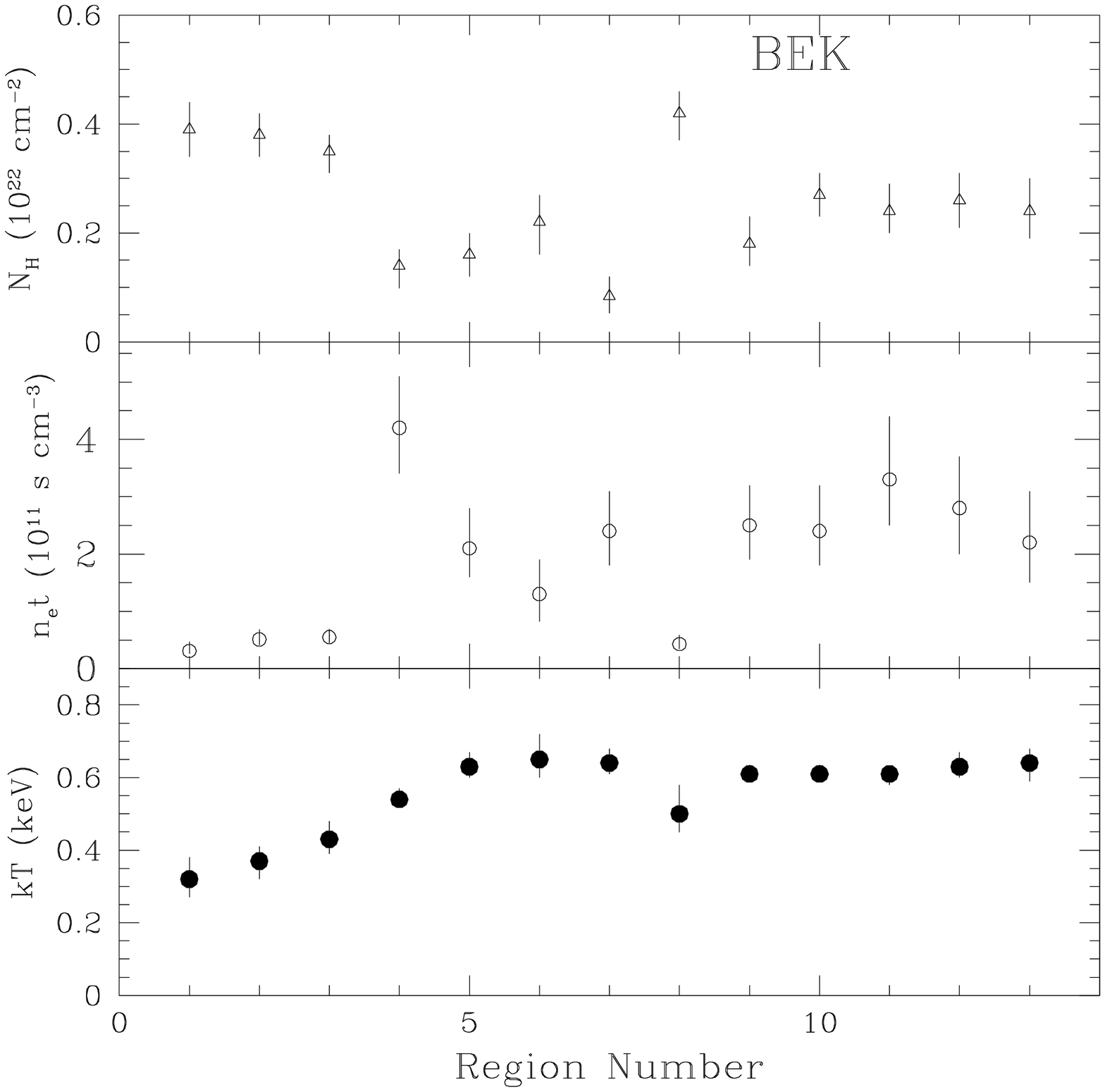}
\caption{(a) Adjacent regions used to take a more detailed look
at the interface between cooler $\sim$ 0.3 keV emission and the hotter
$\sim$ 0.6 keV emission going from the compact knot to the bar. (b)
Fitted column density, ionization age, and temperatures for these
spectra. The regions are numbered successively from east to west.}
\end{figure}

\begin{figure}
\epsscale{1.0}
\label{fig-6}
\plotone{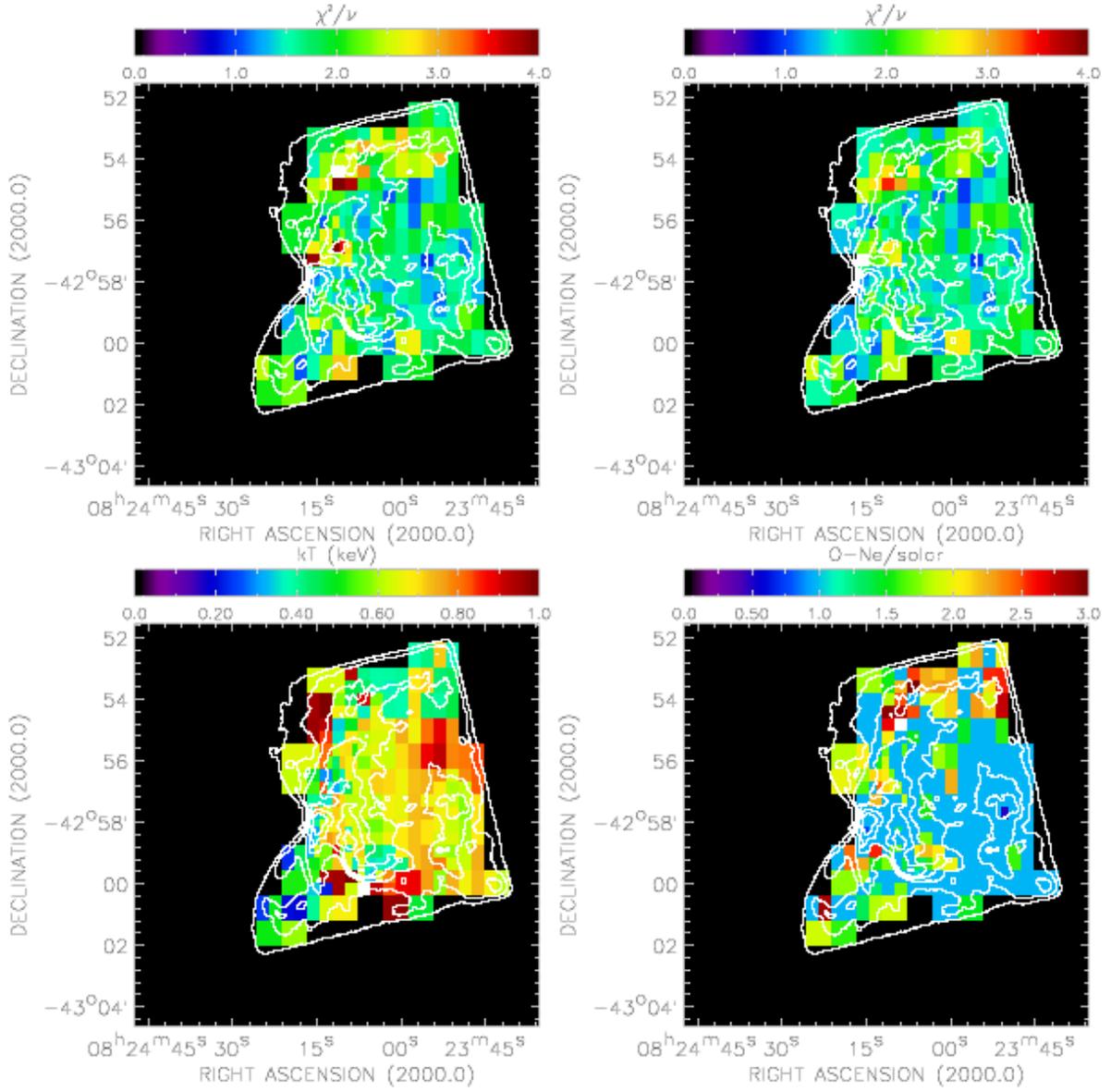}
\caption{(a) Original map of
$\chi^2/\nu$; (b) Map of $\chi^2/\nu$ after allowing O-Ne to be
fitted; (c) Temperature (kT) map in keV when the O-Ne
abundance is allowed to be freely fitted. The revised fit is accepted
for the grid if the improvement in $\chi^2$ satisfies the 0.5\%
probability F-test for infinite degrees of freedom ($\Delta
\chi^2/\chi^2_{r,new} >$ 7.88); (d) Corresponding abundance
map (number abundance relative to the solar values for O and Ne in
their solar abundance ratio).}
\end{figure}

\begin{figure}
\epsscale{0.4}
\label{fig-7}
\plotone{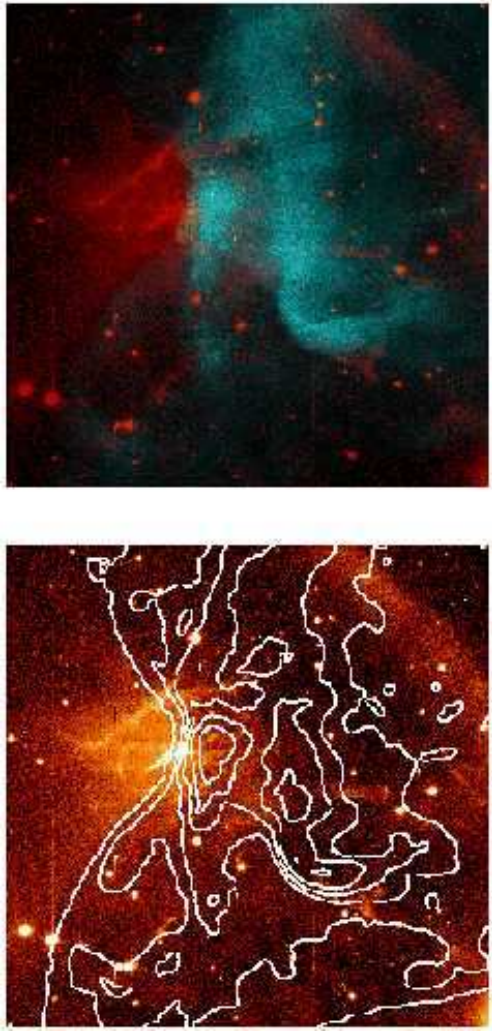}
\caption{The [O III] image of Blair et al. 1995: (a) in red,
with the Chandra X-ray image in blue and (b) with the Chandra
X-ray contours overlaid.}
\end{figure}

\end{document}